# Influence of nonlinear- and saturable-absorption on laser lift-off threshold of oxide/metal structure

Andrius Žemaitis, Mantas Gaidys, Paulius Gečys, Mindaugas Gedvilas*

*Department of Laser Technologies (LTS), Center for Physical Sciences and Technology (FTMC), Savanorių Ave. 231, LT-02300 Vilnius, Lithuania*
*Corresponding author: mgedvilas@ftmc.lt*

**In this work, a new model of effective lift-off threshold of oxide/metal target is presented. The influence of nonlinear processes in the oxide layer on its removal from the metallic samples using picosecond laser was investigated. Nonlinear and saturable absorption in the layer was incorporated into modeling for prediction of effective laser lift-off threshold fluence change with varying peak intensities in *z*-scan type experiment for the first time. The new model coincides well with the experimental results.**

https://doi.org/10.1364/OL.404760

The pulsed laser ablation threshold, laser-induced damage threshold (LIDT), and lift-off threshold fluences are used to describe the minimum laser energy per unit area which is required to remove (ablate) material from the surface. It is well known that the material removal threshold is dependent on many laser processing parameters, such as the number of pulses per spot, irradiation wavelength, [1], pulse duration [2], and polarization type [3]. The laser spot size on the sample and its related peak intensity's influence on the ablation threshold is rarely investigated in scientific literature, especially for metals [4]. Contrary, for dielectric coatings LIDT dependence on the beam spot size is widely known and accepted [5]. A couple of independent scientific groups have developed theoretical models to predict and explain the ablation thresholds of metal dependence on laser beam size [4, 6]. Kautek *et al.* proposed the theory called extended defect model, which explains that the ablation threshold increases when spot size decreases [4], however, Zhang *et al.* predicts completely different behavior [6]. Thus, there is no complete theoretical model found in the scientific literature that would fully explain the variation of the ablation threshold with varying beam sizes and its related peak intensities.

Metals are covered with oxide layers with thicknesses ranging from tens of nanometers to tens of micrometers [7] which usually are semi-opaque semiconductors and transparent insulators depending on the bandgap. Though, incoming laser irradiation firstly interacts with the metal oxides and only afterward with the metals themselves. However, the metal oxides are hard to ablate because they possess high laser ablation thresholds [8] which are by the order of magnitude higher than the metal ablation thresholds [9], and the laser ablation is initiated on an oxide/metal interface during lift-off process [10]. The nonlinear- and saturable-absorption [11] are commonly observed when metal oxide films are exposed to intense light with intensities ranging from MW/cm$^2$ to TW/cm$^2$. Therefore, those oxide layers covering the metal affects the fraction of the incoming laser irradiation dose that reaches the target substrate depending on the intensity applied. The ablation threshold dependencies on pulse duration and its related peak intensities are well known for metal and dielectrics [12], but have never been investigated for an oxide/metal system. The laser lift-off threshold of the metal covered by an oxide is influenced by the nonlinear processes in the oxide layer and has to be taken into account. Thus, the laser lift-off threshold of an oxide/metal depends on the laser intensity applied and can be called the effective lift-off threshold fluence. Moreover, interference related field enhancement effects influence the oxide layer ablation [13], though, it will not be included in our work, since the transfer-matrix method is required for absorbed energy calculations, and nonlinear processes cannot be applied for this method. To the best of our knowledge, the influence of nonlinear- and saturable-absorption in the metal oxide layer on the effective laser lift-off threshold of an oxide/metal structure has never been investigated in the scientific literature. Therefore, novel experimental and theoretical research on the effective laser lift-off threshold change with varying peak intensities in *z*-scan experiments has been created in this work for the first time.

Here we propose the nonlinear- and saturable-absorption in the oxide layer on the metal substrate as the two most influencing factors for the effective laser lift-off threshold dependence on the different peak intensities in *z*-scan experiments. The existing analytical normalized transmission equation of the semi-transparent thin layer in open aperture *z*-scan was incorporated into the effective laser lift-off threshold dependence on the sample position in the newly presented model. The *z*-scan type picosecond laser ablation experiment was conducted, and lift-off thresholds were measured at different sample positions, which caused different beam peak intensities on the processed oxidized copper target material. The numerical calculations using the new proposed model equation coincides well with the experimental results of effective laser lift-off threshold dependence on the z-scan position. The copper (Cu) sample with cuprous oxide ($Cu_2O$) layer was used in the laser ablation experiments. The copper (CW004A, Ekstremalė) target had mirror-like-finish with the surface roughness $R_a$ < 0.1 μm measured by a stylus profiler (Dektak 150, Veeco). The picosecond laser irradiation source (Atlantic, Ekspla) with a pulse duration $\tau$ = 10 ps, wavelength $\lambda$ = 1064 nm, pulse energies up to $E_p$ = 142 μJ, repetition rate $f_p$ = 100 kHz, Gaussian beam with a quality parameter of $M^2$ = 1.06 measured in our previous work [14], and a number of laser pulses per spot $N$ = 1 was used in ablation experiments. The laser-treated oxide/metal surface topography was studied by scanning electron microscope (SEM) (JSM-6490LV, JEOL) (Fig. 1(a)) and optical profiler (S neox, Sensofar) ((Fig. 1(b)).

The laser-irradiated area consisted of two distinct areas, laser ablation of pristine copper and lift-off of the oxide layer ((Fig. 1(a)). The thickness of the oxide layer measured from the height profile was $L$ = 1.0 ± 0.1 μm ((Fig. 1(b)). The fluence threshold was measured by the ablated crater diameter $D$ dependence on the pulse energy $E_p$ [15]. Taking into account that a laser beam has the Gaussian transverse intensity distribution, the expression for the crater diameter squared $D^2$ and the laser pulse energy $E_p$ can be written as [15]:

$$D^2 = 2w^2 \ln\left(2E_p / \left(\pi w^2 F_{th}\right)\right), \quad (1)$$

where $w$ is the Gaussian beam radius on the sample, $E_p$ is the pulse energy, and $F_{th}$ is the threshold fluence. The line extrapolation of $D^2(E_p)$ to the $D^2$ = 0 μm² gives the threshold pulse energy $E_{pth}$, the function slope is related to the beam radius $w$, and the ablation threshold than can be evaluated by $F_{th} = 2E_{pth}/(\pi w^2)$. The optical microscope (Eclipse LV100, Nikon) was used for the measurements of the ablated diameters of pristine copper and diameters of the removed oxide layer of $Cu_2O$/Cu samples (Fig. 1(c)). Threshold values of $F_{th}$ = 0.57 ± 0.03 J/cm² and 2.6 ± 0.3 J/cm² were retrieved from line fits by Eq. (1) for oxide lift-off and copper ablation, respectively (Fig. 1(c)). The literature value of the lift-off threshold for copper oxide removal from the copper sample is $F_{th}$ = 0.62 J/cm² ($\tau$ = 12 ps, $\lambda$ = 1064 nm, $f_p$ = 100 kHz, $w$ = 17 μm, $N$ = 1) [10]. The literature values of ablation threshold of copper by ($\tau$ = 10 ps, $\lambda$ = 1064 nm, $f_p$ = 50 kHz, $N$ = 1) laser at different beam radiuses $w$ are: $F_{th}$ = 2.0 J/cm² ($w$ = 41.8 μm) [14]; $F_{th}$ = 1.73 J/cm² ($w$ = 20 μm) [16]; $F_{th}$ = 0.95 J/cm² ($w$ = 17.5 μm) [17]. The literature threshold fluence values coincided well with our current work.

The light attenuation propagating through a linearly and nonlinearly absorbing material can be described as [18]:

$$dI/dz = -\alpha I - \beta I^2, \quad (2)$$

where $I = I(z)$ is the peak intensity within the sample material, depending on the propagation distance $z$ in the sample, $\alpha$ is the linear absorption coefficient, and $\beta$ is the nonlinear absorption coefficient. The normalized transmission through the sample can be extracted from the solution of Eq. (2) and expressed as [19]:

$$T = e^{-\alpha L}/\left(\beta I L_{eff} + 1\right), \quad (3)$$

where $I$ is the incident intensity, $L$ is the physical sample thickness, and $L_{eff}$ is the effective thickness of the sample [19]:

$$L_{eff} = \left(1 - e^{-\alpha L}\right)/\alpha. \quad (4)$$

Because of the linear absorption saturation behavior, the linear absorption coefficient $\alpha = \alpha(I)$ depends on the peak intensity $I$, and for homogenous saturation can be expressed as [20]:

$$\alpha(I) = \alpha_0 / \left(1 + (I/I_s)^2\right), \quad (5)$$

where $\alpha_0$ is the low-intensity absorption coefficient and $I_s$ is the saturation intensity. The Gaussian beam intensity in the z-scan type experiment can be expressed as:

$$I = I_0 / \left(1 + (z/z_R)^2\right), \quad (6)$$

where $I_0 = 2E_p/(\pi w_0^2 \tau)$ is the peak pulse intensity in the center of the Gaussian beam at a focus position, $w_0$ is the focused beam radius at $1/e^2$ level, $\tau$ is the pulse duration, $z_R = \pi w_0^2/(\lambda M^2)$ is the Rayleigh length, $\lambda$ is the wavelength of irradiation, and $M^2$ is the quality parameter of the Gaussian beam.

The direct bandgap for $Cu_2O$ is 2.1 eV [21], which is larger than the photon energy 1.17 eV at the wavelength of 1064 nm. The linear absorption coefficient of $Cu_2O$ is $\alpha$ = 5.2 × 10³ cm⁻¹ [22]. Therefore, the transmission of linearly absorbing oxide layer at low peak intensities in the GW/cm² range evaluated by Eq. (3) is $T_0 = \exp(-\alpha_0 L) \approx 60\%$, and the layer can be assumed as semi-opaque. The saturation intensity of the linear absorption coefficient for $Cu_2O$ is $I_s \sim 0.1$ TW/cm² [23], thus, the semi-opaque oxide layer suddenly becomes fully transparent, when peak intensities exceed the saturation intensity. However, by further increasing intensity the nonlinear absorption becomes dominant and the oxide layer changes to opaque again. The nonlinear absorption coefficients of $Cu_2O$ is $\beta$ = 4.3 × 10⁻⁹ cm/W [22]. Thus, a focused picosecond laser beam with high peak intensities up to 3 TW/cm² obtained in our experimental conditions induces nonlinear absorption in the transparent oxide layer on top of the copper target. The nonlinearly absorbing oxide layer transmits only a part of the incoming laser beam to the metal substrate and has to be taken into account when predicting effective ablation characteristics of an oxide/metal structure. The Eqs. (3), (4), (5), and (6) have been numerically solved and normalized and the $T/T_0$ transmission of the $Cu_2O$ layer was calculated (Fig. 2(a)).

At low peak pulse intensities of $I_0 \approx 0.05$ TW/cm² that are smaller than the saturation intensity $I_s \sim 0.1$ TW/cm² just linear absorption is observed with no saturation and very small nonlinear absorption in the focal position $z \ll \pm z_R$, thus, almost constant normalized transmission $T/T_0 \approx 1$ in all sample $z$ positions is observed (Fig. 2(b)). With the peak intensity values of ~ 0.2 TW/cm$^2$ exceeding the saturation intensity, the oxide layer becomes nearly fully transparent $T/T_0 \approx 1.7$ with a sample position close to the focal point $z \ll \pm z_R$ (Fig. 2(c)). When peak intensity reaches the value of ~ 0.5 TW/cm$^2$ the nonlinear absorption equals the linear intensity saturation, thus the normalized transmission in the oxide layer reaches a value close to $T/T_0 \approx 1$ in the focal region $z \approx 0$ mm (Fig. 2(d)). However, the nonlinear absorption is very sensitive to the peak intensity and occurs only in fine focus $z \ll \pm z_R$. Out of the focus in $z \approx \pm 1$ mm the Cu$_2$O layer is still transparent with a transmission value of $T/T_0 \approx 1.7$ (Fig. 2(d)). When peak pulse intensity reaches a high value of ~ 0.9 TW/cm$^2$ the nonlinear absorption exceeds the absorption saturation and normalized transmission drops to the value of $T/T_0 \approx 0.4$ in the focal point. Therefore, the Cu$_2$O layer becomes opaque in the focal range of the beam. However, the defocused beam in the Rayleigh range $z \approx \pm z_R$ is still transparent because of the saturation of linear absorption with normalized transparency of $T/T_0 \approx 1.7$ (Fig. 2(e)).

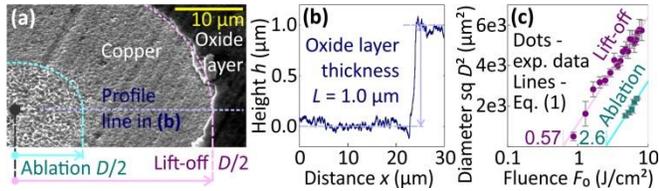

**Fig. 1.** (a) SEM micrograph, (b) profile, and (c) diameter squared $D^2$ versus peak laser fluence $F_0$ of ablated copper and removed oxide layer from copper substrate by lift-off technique. SEM image was taken at tilt angle of 45°, pulse duration $\tau$ = 10 ps, wavelength $\lambda$ = 1064 nm, repetition rate $f_p$ = 100 kHz, number of pulses $N$ = 1, pulse energy $E_p$ = 51 μJ, beam radius $w$ = 33.2 ± 0.7 μm, peak laser fluence $F_0$ = 3.0 J/cm², peak intensity $I_0$ = 0.30 TW/cm², ablated diameter of pristine copper $D$ = 16.9 ± 1.8 μm, lift-off diameter of oxide layer $D$ = 60.4 ± 1.6 μm (a). The oxide layer thickness $L$ = 1.0 ± 0.1 μm evaluated from height profile (b). The oxide lift-off threshold fluence 0.57 ± 0.03 J/cm² and copper ablation threshold 2.6 ± 0.3 J/cm² were retrieved from line fits of experimental data points by Eq. (1) in (c).

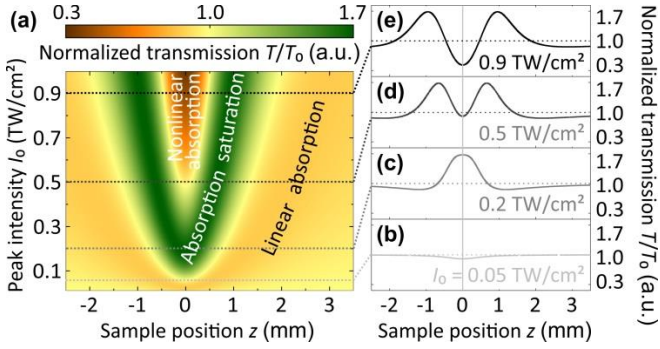

**Fig. 2.** (a) Calculated normalized transmission $T/T_0$ (color scale) versus sample position $z$ (bottom axis) and peak pulse intensities $I_0$ (left axis) for Cu$_2$O layer. The three different absorption regions are indicated: yellow – linear absorption; green – linear absorption saturation; brown – nonlinear absorption. Profiles of normalized transmission $T/T_0$ (right axis) dependence on the sample position $z$ (bottom axis) at different peak intensities $I_0$: (b) 0.05 TW/cm$^2$; (c) 0.2 TW/cm$^2$; (d) 0.5 TW/cm$^2$; (e) 0.9 TW/cm$^2$.

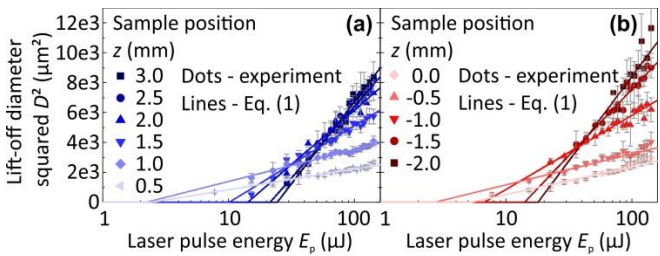

**Fig. 3.** Lift-off diameter squared $D^2$ of oxide layer removed from copper dependence on the laser pulse energy $E_p$ at various sample position $z$ values ranging from 0.5 mm to 3.0 mm, step 0.5 mm (a) and from 0.0 mm to -2.0 mm, step -0.5 mm (b). Solid dots represent the experimental data points, error bars are the statistical deviation from average crater diameter from 10 dimples ablated using the same pulse energy, and solid lines are fits by Eq. (1). Pulse duration $\tau$ = 10 ps, laser wavelength $\lambda$ = 1064 nm, repetition rate $f_p$ = 100 kHz, and number of pulses per spot $N$ = 1.

The $z$-scan type laser lift-off threshold evaluation experiments with controlled laser peak pulse intensities have been conducted in this work. The peak pulse intensities $I_0$ were ranging from 20 GW/cm$^2$ to 3.0 TW/cm$^2$. The Rayleigh length of the focused beam was of $z_R \approx 0.83$ mm. The squared lift-off diameters of an oxide layer from copper at different positions $z$ are depicted in (Fig. 3(a) and (b)).

Pulse energies of 20 different values have been used at each of the 11 fixed $z$-scan positions with fixed laser spot radius on the sample. Every test has been repeated 10 times to get the statistical deviation errors bars. Experimental data points of ablated crater diameters were fitted using Eq. (1) and laser spot radiuses at different sample positions $z$ were retrieved as fitting parameters and its errors. The laser spot radius values on the sample ranged from $w_0 = 17.2 \pm 0.3$ µm in the focal region to $w = 48.6 \pm 1.2$ µm out of focus. The fit of experimental data points by Eq. (1) in Fig. 3(a) and (b) also provided information about the effective lift-off threshold fluence of oxide/copper structure at various positions $z$ of the sample (Fig. 4(a)).

It was found out that the effective lift-off threshold is highest near the beam waist $z = 0$ mm, then we have seen a sharp drop of effective lift-off threshold to a minimum value at $z = \pm 1$ mm, and afterward it increased again (Fig. 4(a)). A similar drop of the laser-ablated volume per pulse in $z$-scan ablation at the sample position of $z/z_R \approx 0$ has been observed by Chen *et al.* [24], however, the drop of ablation rate was not attributed to the increase of threshold fluence in their research. Thus, in this work, we have proposed a new model for the effective lift-off threshold fluence which considers the attenuation of the incoming laser irradiation by the oxide layer on the metal substrate.

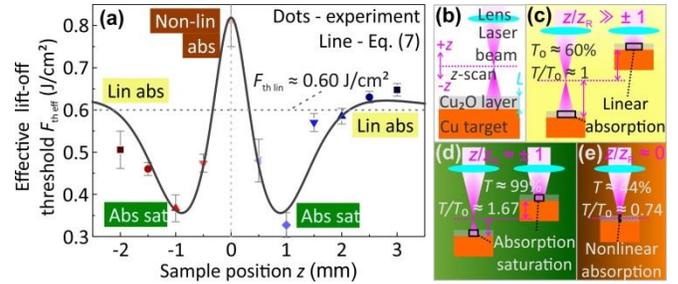

**Fig. 4.** (a) Effective lift-off threshold $F_{th\,eff}$ of oxide/copper dependence on sample $z$ position. Solid dots and error bars represent the effective lift-off threshold values retrieved from the fitting parameters by Eq. (1) in Fig. 3(a) and (b), the solid grey line is the fit of experimental data points by Eq. (7). Pulse duration $\tau$ = 10 ps, wavelength $\lambda$ = 1064 nm, repetition rate $f_p$ = 100 kHz, and a number of pulses per spot $N$ = 1. (b) Principal visual scheme of the proposed model with semi-transparent Cu$_2$O layer on top of Cu sample, which partially attenuates the focused picosecond laser beam in $z$-scan laser ablation experiment with indicated incoming Gaussian beam, focusing lens, and variable sample positions ranging from top $+z$ to bottom $-z$ in respect with the beam focus and its related controlled peak intensities. The copper substrate is indicated with orange color, the semi-opaque cuprous oxide layer in grey color. The different Cu$_2$O layer transmission cases are indicated depending on peak pulse intensities: (c) out of focus with sample positions $z/z_R \gg \pm 1$, low intensities of the defocused beam and linear absorption with transmission value $T_0 \approx 60\%$; (d) sample position $z/z_R \approx \pm 1$ in the saturable absorption intensity region with transmission value $T \approx 99\%$; (e) sample position $z/z_R \approx 0$ in the focus of the beam and high peak intensities, nonlinear absorption exceeds absorption saturation in the oxide layer and transmission drop to $T \approx 44\%$.

In our proposed ablation model, the laser irradiation is attenuated by the semi-transparent oxide layer, and only a fraction of the incoming laser beam reaches the copper target. The principal visual scheme of our threshold model of a semi-transparent $Cu_2O$ layer with the thickness of $L = 1.0$ µm on the Cu substrate is depicted in Fig. 4(b). The Gaussian picosecond laser beam is focused by the objective lens and the peak pulse intensity is varied on the target material by controlling the $z$-scan position of the sample. The three different absorbing scenarios depending on the peak pulse intensities applied are: linear absorption, a saturation of linear absorption, and nonlinear absorption depicted in Fig. 4(c), (d), and (e), respectively. By having the sample out of the focal position, moderate values of effective lift-off threshold of $F_{th\,lin} \approx 0.60$ J/cm$^2$ are observed (Fig. 4(a)). This is explained by our proposed model when highly defocused beam $z/z_R \gg \pm 1$ with low peak intensities is applied on the sample, only linear absorption is observed in the attenuating $Cu_2O$ layer (Fig. 4(c)), thus, just a fraction $T_0 \approx 60\%$ of the incoming laser beam interacts with the copper in the oxide/metal interface, and moderate energy densities are needed to ablate the copper target material. However, when laser intensities reach the saturable absorption, the oxide layer suddenly becomes transparent and almost all incoming laser irradiation reaches the pristine copper. Thus, the drop of effective lift-off threshold to $\sim 0.36$ J/cm$^2$ is observed (Fig. 4(a)). It is graphically depicted in our proposed model with moderate intensities above the saturation intensity at sample position $z/z_R \approx \pm 1$ and moderate peak intensities with the normalized transmission of the oxide layer of $T/T_0 \approx 1.67$ (Fig. 4(d)). Thus, the major part of the incoming laser irradiation $T \approx 99\%$ is transmitted through the oxide layer. When peak pulse intensities in the focal range reach the TW/cm$^2$ values, the nonlinear absorption becomes dominant and exceeds the absorption saturation, consequently, only a fraction of the incoming laser beam reaches the copper, and the effective lift-off threshold increases to the value of 0.81 J/cm$^2$ (Fig. 4(a)). It is explained with our model that at the focal position $z/z_R \approx 0$ the beam is focused and the highest peak intensities are achieved, therefore, the nonlinear absorption is reached and the oxide layer becomes opaque $T/T_0 \approx 0.74$ (Fig. 4(e)). Only a small fraction $T \approx 44\%$ of the incoming laser beam goes through the $Cu_2O$ layer and reaches the Cu substrate.

Therefore, taking into account linear absorption, absorption saturation, and nonlinear absorption in the model it can be assumed that the effective lift-off threshold of oxide/metal is the product of the ablation threshold in linear absorption regime $F_{th\,lin}$ at low peak intensities and the inverse normalized transitivity $T_0/T$ of cuprous oxide layer defined by Eq. (3). The effective lift-off threshold can be expressed as follows:

$$F_{th\,eff} = F_{th\,lin}\left(T_0/T\right). \qquad (7)$$

The effective lift-off threshold $F_{th\,eff} = F_{th\,eff}(z)$ defined by Eq. (7) depends on the sample position $z$ in the $z$-scan type laser ablation experiments. The model fit is in good coincidence with the experiment (Fig. 4(a)). The new model predicts the fall of the effective ablation threshold to minimum values of $F_{th\,eff} \sim 0.36$ J/cm$^2$ near the sample position of $z \approx \pm 1$ mm. Also, the model predicts a steep rise of ablations threshold value up to $F_{th\,eff} \sim 0.81$ J/cm$^2$ in the focus of the beam. The moderate values of effective ablation threshold $F_{th\,eff}$ close to $F_{th\,lin} \approx 0.60$ J/cm$^2$ out of the focus are also predicted.

To conclude, the nonlinear absorption and saturable linear absorption have been incorporated in a theory of effective laser lift-off threshold of oxide/metal structure dependence on a $z$-scan position related peak intensities for the first time. The attenuation of the incoming laser beam by a copper oxide layer on a copper target was included in the ablation model. The new model has a good agreement with the experimental data of the $z$-scan type effective lift-off threshold of the cuprous oxide/copper target. Our proposed model opens new insights on the influence of the surface oxide layer on the effective laser lift-off threshold of oxidized metallic samples.